\documentclass[11pt,a4paper]{article}

\usepackage[T1]{fontenc}
\usepackage[utf8]{inputenc}
\usepackage{lmodern}

\usepackage[margin=1in]{geometry}
\usepackage{setspace}
\usepackage{amsmath,amssymb,amsthm}

\usepackage{graphicx}
\usepackage{booktabs}

\usepackage{algorithm}
\usepackage{algpseudocode}

\usepackage{tikz}
\usepackage{eso-pic}
\newcommand\WatermarkText{%
  \begin{tikzpicture}[remember picture,overlay]
    \node [rotate=45,scale=8,text opacity=0.1] 
      at (current page.center) {DRAFT};
  \end{tikzpicture}%
}
\AddToShipoutPictureBG{\WatermarkText}

\usepackage[
  backend=biber,
  style=authoryear,
  citestyle=authoryear, 
  maxcitenames=2,
  maxbibnames=99
]{biblatex}
\usepackage[colorlinks=true,allcolors=blue]{hyperref}

\usepackage{authblk} 

\begin{document}

\title{Transdimensional Data Assimilation for dynamic model selection problems}

\author[1]{Márk Somogyvári}
\author[2]{Sebastian Reich}

\affil[1]{HU Berlin, IRI THESys, \texttt{mark.somogyvari@hu-berlin.de}}
\affil[2]{University of Potsdam, Institute of Mathematics, \texttt{sebastian.reich@uni-potsdam.de}}

\date{\today}
\maketitle

\begin{abstract}
In this paper we combine the non-linear filtering capabilities of particle filters with the transdimensional inference of the reversible-jump Markov chain Monte Carlo method for a data assimilation methodology over dynamic problems with variable dimensionality. By using transdimensional MCMC steps for the rejuvenation of the particle filter, the algorithm could change the number of state space parameters on the fly and can be applied for transdimensional data assimilation purposes. Classic inversion methodologies use pre-defined models, and only changes the individual parameter values during interpretation. This is often not feasible when the optimal model parametrization is not known a priori or when the model resolution needs to change with time. The proposed transdimensional particle filter algorithm, combines the advantages of particle filters and the transdimensional MCMC methods, and provides an easily implementable data assimilation algorithm that could tackle such problems. The methodology could also improve the computational efficiency of particle filters as it could inherently optimize the model complexity in a data-driven way.
We demonstrate the capabilities of the enhanced algorithm on two simple model examples.
\end{abstract}

\section{Introduction}
Classic data assimilation methods such as the Kalman-filter \parencite{Kalman1961}, or the ensemble Kalman filter \parencite{Evensen2009} use the evolution of covariances within the state vector to make predictions on the state space model. While the variants of these data assimilation techniques became extremely popular, their reliance on linear parameter relations and Gaussian probability distributions still poses a challenge for their application in practical cases \parencite{VanLeeuwen2019}. In contrast, particle filters do not assume linearity or Gaussian distributions \parencite{Doucet2001,Gordon1993}. Particle filters are sequential Monte Carlo methods (often referred to as SMC) that use a random set of particles (an ensemble) to approximate the probability density function of the state space vector given the observations. The core concept of particle filters is, when the ensemble size is large enough, the particle set provides an equivalent representation of the target distribution \parencite{Gordon1993}.

Data assimilation methods are computationally expensive in general, particle filters especially. Particle filters use random perturbations as the underlying drivers for the convergence to the target distribution, similarly to MCMC algorithms. To converge to the target distribution, this requires significant number of iterations and properly chosen prior distributions \parencite{Vetra-Carvalho2018}. These problems explode with the increase of model parameter numbers (model dimensions), making the methodology challenging to use for complex models \parencite{VanLeeuwen2019}.

In high dimensions particle filters are also more prone to weight degeneracy and particle impoverishment (also known as filter collapse) \parencite{Morzfeld2017,Surace2019}. High dimensional state-space vectors require high dimensional observations, which leads to big differences in the individual particle weights. Because of this, most particles lose their significance during the data assimilation process and the effective sample size rapidly goes down to a few or only one particle. Using more particles could mitigate this problem, but the increase needs to be exponential with the observation vector size \parencite{Snyder2008}, which leads to an exponential increase in computational costs.

To overcome the challenges of high dimensional problems, particle filter variants use optimal transport, localization techniques, adaptive resampling or hybridization with ensemble Kalman filters \parencite{VanLeeuwen2019}. In other applications, the high dimensional problems are simplified via the transformation of variables, or by the elimination of the non-relevant parameters \parencite{VanLeeuwen2009a,Kantas2015}. Elements of the state space model have different importance for describing the modelled system, and these discrepancies could change over time. The model could be insensitive to a state space parameter at one point in time, but the same parameter could become relevant again in a later time. In these cases, it would be advantageous to temporarily remove the insensitive model elements hence reducing the complexity of the model.

Classic data assimilation and inversion methods deal with problems were the underlying model is defined as a fixed number of parameters \parencite{Sambridge2006}. There are situations however where this traditional approach is not feasible. In practical applications often not enough information is available prior the interpretation and the model that gives the best fit to the observed data needs to be selected. Physical problems with spatially variable resolution are notable examples, like seismic tomography \parencite{Sambridge2012}.

Transdimensional inversion techniques, such as the Metropolis-Hastings-Green algorithm tests different parameter numbers to select the ideal model for the investigated problem \parencite{Green1995}. By allowing jumps between different model dimensions, this generalized version of MCMC algorithms could explore parameter spaces of different dimensions, while maintaining the stationarity of the used Markov chains \parencite{Brooks2011}. The Metropolis-Hastings-Green algorithm became a standard tool today for transdimensional inversion, but no similar method has been proposed so far for problems with a temporal dynamic aspect.

A trivial solution to deal with model selection problems in a data assimilation framework is to apply a transdimensional inversion prior the data assimilation on the initial state space model. However this strategy only works when the required model parameterization is unknown but static. This research is mainly motivated by a different type of problems, when the optimal model parameterization is changing over time. In this case the transdimensional inversion needed to be integrated into the data assimilation algorithm directly. Object tracking with changing number of objects or adaptive discretization problems could be notable examples. As an additional benefit, transdimensional data assimilation could also reduce model dimensions of high dimensional models by omitting non-relevant parameters \parencite{Jimenez2016}. Therefore it could provide computationally efficient alternatives for high dimensional data assimilation problems.
\parencite{larocque2002particle} used such methodology for multi-object tracking problem, but the topic did not gain much attention since. In the following we follow a similar approach to their work with a methodology that includes transdimensional MCMC updates into standard particle filter algorithms. As many particle filter approaches already use MCMC simulations for avoiding filter impoverishment, with the modification of these chains the algorithms could be extended over dimensions.
Our paper is structured as follows. In the next section we present the standard bootstrap filter algorithm, then the transdimensional particle filter for transdimensional data assimilation. In section 3, we demonstrate the applicability of the method on a signal processing example. In section 4 we present an object tracking problem with increased complexity. We conclude our paper with some general remarks about the extended algorithm, and with outlook to further improvements and to potential applications.

\section{Methodology}
In this section we present the algorithm for the transdimensionally-extended particle filter. As a basis we use the standard bootstrap filter, one of the simplest variants of particle filters \parencite{Gordon1993}. This sequential method uses a series of importance sampling to map the evolution of the probability density function (PDF) of an observed dynamic system.
At any time, the posterior PDF of the problem can be calculated from the prior and the likelihoods using Bayes' theorem \parencite{Doucet2001a}.
  \begin{equation}\label{prior_and_data}
        \pi_t \left( \theta_{t} | \xi_{1:t} \right) = \frac{\pi_{t} \left( \xi_{1:t} | \theta_{t}\right) \pi_t\left( \theta_{t} \right)}{\int \pi_{t} \left( \xi_{1:t} | \theta_{t}\right)\pi_t\left( \theta_{t} \right) d\theta_{t}}
  \end{equation}

where $\pi_{t} \left( \xi_{1:t} | \theta_{t}\right)$ is the data likelihood, $\pi_t\left( \theta_{t} \right)$ is the prior distribution and the denominator is the evidence, a normalization of the PDF given the observations.\\
The analytical calculation of eq. \ref{prior_and_data} is only possible for relatively small and restricted problems. To estimate more complex PDFs particle filters use particles, large sets of model realizations (also known as the ensemble). Together the particles provide an approximation of the target probability distribution of the system state ($\pi_t$) at every discrete timestep ($t$):

  \begin{equation}\label{particle_density}
        \pi_t \left( \theta_t | \xi_{1:t} \right)\approx \sum_{i=1}^{N}{w_t^i \delta_{\theta}\left( \theta_t^i-\theta_t \right)}
  \end{equation}

where $\theta_t$ is the state space vector at time $t$, $\xi$ is the observation vector, $w_t^i$ are the particle weights and $\delta_{\theta}\left(\theta_t^i-\theta_t \right)$ denotes point samples from the state space \cite{Fearnhead2017}.

In the case of transdimensional problems the formulation of eq. \ref{prior_and_data} remains formally unchanged:
  \begin{equation}\label{prior_and_data2}
        \pi_t \left( \theta_{t}^* | \xi_{1:t} \right) = \frac{\pi_{t} \left( \xi_{1:t} | \theta_{t}^*\right) \pi_t\left( \theta_{t}^* \right)}{\int \pi_{t} \left( \xi_{1:t} | \theta_{t}^*\right)\pi_t\left( \theta_{t}^* \right) d\theta_{t}^*}
  \end{equation}
  The difference for this case is that $\theta_{t}^*$ and $\theta_{t-1}^*$ are not necessarily of the same dimensions for any given time. This applies to the posterior PDF as well, meaning that the standard recursive formulations for eq. \ref{prior_and_data2} are not applicable anymore. This poses a challenge for applying the particle approximation too. In a transdimensional context the dimensionality of the particles have to be unrestricted, they need to be allowed to change dimensions whilst still being part of the same probability distribution.

\subsection{Bootstrap particle filter}
We the here presented simple bootstrap filter as a basis for the transdimensional data assimilation. For initialization at $t=0$, the initial particle set of $N$ particles is generated (drawn from a prior $\pi_0$):
\begin{equation} \label{particles1}
	\{\theta_t^i\} = \theta_t^0,\theta_t^1... \theta_t^{N-1}
\end{equation} 

This initial ensemble could be completely uncalibrated, but in practice it is usually generated after the available prior information. When past observations are available, the initial particles could also be the result of a previous inversion. The particle filter itself is an iterative algorithm constructed from the following steps:

\begin{algorithm}
\caption{Bootstrap filter}
\label{bootstrap}
\begin{algorithmic}[1]
\State reweight  \quad $w_t^i = \frac{\pi_t \left( \xi_t | \theta_t^i \right)}{\sum \pi_t \left( \xi_t | \theta_t^i \right)}$
\State resample \quad $\left( \theta_t^i,w_t^i \right) \rightarrow \left( \theta_t^{i,r} , \frac{1}{N} \right)$
\State rejuvenate \quad $\theta_t^{i,m} \sim K_t \left( \theta_t^{i,r}, \cdot \right)$
\State propagate \quad $t = t + 1 \quad \theta_t^{i,m} \rightarrow \theta_{t+1}^i$
\end{algorithmic}
\end{algorithm}

Every iteration starts with an importance sampling step, a reweighting \cite{Geweke1989}. Here a corresponding weight is calculated for each particle, representing its probability: 
\begin{equation}
	w_t^i = \frac{\pi_t \left( \xi_t | \theta_t^i \right)}{\sum \pi_t \left( \xi_t | \theta_t^i \right)}
\end{equation}

The weighted particles could be used as an approximator of the target distribution.  The next step is the resampling, where all the existing particles are replaced with particles of weights of one:

\begin{equation}
	\left(\theta_t^i,w_t^i \right) \rightarrow \left( \theta_t^{i,r},\frac{1}{N} \right)
\end{equation}

Particles with large weights before are replaced with multiple new particles, while particles with small weights are replaced with single new ones or with none. \\
Iterating only the reweight-resampling steps over time would lead to filter degeneracy \cite{Chopin2002}. Each reweighting step introduces more variability, which will eventually lead to very few particles with significant weights. The subsequent resampling removes the low weight particles (hence  saving computational time by removing the need of simulating unlikely models), but it will only generates new ones from a few high weight particles. The resampled particle set will be strongly correlated, impoverished \cite{Li2013}.\\
The classic solution for particle impoverishment is the introduction of a rejuvenation step to the particle filter. To enrich the particle set, short series of Markov chain Monte Carlo (MCMC) simulations are run on each particle \cite{Li2014a}. These MCMC chains move the particles around in the parameter space, hence obtaining better coverage over the target distribution. To maintain the distribution of the particle system, stationary Markov kernels are used within the MCMC chains:

\begin{equation}
	\theta_t^{i,m} \sim K_t \left( \theta_t^{i,r},\cdot \right)
\end{equation}

Stationarity of MCMC algorithms are usually obtained by using reversible chains \cite{Geyer2011}. Reversible MCMC chains employ Markov kernels ($K(x,\cdot)$) where:

\begin{equation} \label{stationarity}
	\pi(dx)K(x,dy)=\pi(dy)K(y,dx)
\end{equation}

meaning that the probability of any choosen forward step in the Markov chain equals with the probability of the same step reversed. If this expression holds true, $\pi$ is a stationary distribution of the Markov chain. For rejuvenation any type of Markov kernel could be used, given it fulfills the criteria of stationarity.\\
Because this step is significantly more cost intensive than the previous ones (as it requires multiple runs of the observation model at each particle), it is not called at every iteration. Instead it is just called at every n-th timestep, or when the impoverishment of the particle set falls below a critical level.
The final step of one particle filter iteration is the propagation of the particles in time according to the underlying dynamic model. Hence a forecast of the state of the dynamic system at the next timestep is created.

\subsection{Transdimensionally-extended particle filter}
When dealing with a transdimensional problem the dimensionality of the PDF of the state space changes over time. This could be for example due to the appearance of new objects in an object tracking example, but it could also happen with a model selection problem, where a model with more parameters is required to properly infer the state space in later timesteps. Standard data assimilation methods not perform very well in such situations, as they are designed to work in fixed dimensionalities.
\cite{Green1995} introduced the reversible-jump MCMC (rjMCMC) and showed, that reversibility (hence stationarity) can be ensured even when the chain jumps between dimensions. The rjMCMC algorithm uses the Metropolis-Hastings-Green (MHG) kernel for transdimensional jumps between state $x$ and $y$:

\begin{equation}
	K(x,y) = 
	\begin{cases}
	accept \quad \alpha(x,y)=min(1,r(x,y))\\
	reject \quad 1-\alpha(x,y)
	\end{cases}
\end{equation}

where $\alpha$ is the Green-ratio, calculated as:

\begin{equation}
	\alpha(x,y)= \frac{p(y)p(y|\xi)q(x|y)}{p(x)p(x|\xi)q(y|x)} |J|
\end{equation}

In this equation $\frac{p(y)p(y|\xi)}{p(x)p(x|\xi)}$ is the ratio of the posteriors, $\frac{q(x|y)}{q(y|x)}$ is the proposal ratio and $|J|$ is the so-called Jacobian matrix.
The MHG kernel keeps the MCMC chains within the same meta-distribution, keeping the stationarity at transdimensional jumps. Hence, the rjMCMC method is a good solution for using Bayesian inference with model selection problems.
Because of their stationarity, rjMCMC chains can be integrated into the bootstrap filter algorithm as part of the rejuvenation step. Hence for the transdimensional extension, other parts of the particle filter algorithm do not need any modification. Therefore a transdimensionally-extended particle filter algorithm could be written similar as the bootstrap filter:

\begin{algorithm}
\caption{Transdimensional bootstrap filter}
\label{TDfilter}
\begin{algorithmic}[1]
\State reweight  \quad $w_t^i = \frac{\pi_t \left( \xi_t | \theta_t^i \right)}{\sum \pi_t \left( \xi_t | \theta_t^i \right)}$
\State resample \quad $\left( \theta_t^i,w_t^i \right) \rightarrow \left( \theta_t^{i,r} , \frac{1}{N} \right)$
\State rejuvenate \quad $\theta_t^{i,m} \sim K_t^* \left( \theta_t^{i,r}, \cdot \right)$ - MHG kernel
\State propagate \quad $t = t + 1 \quad \theta_t^{i,m} \rightarrow \theta_{t+1}^i$
\end{algorithmic}
\end{algorithm}

Every step of the modified algorithm is the same, except for the modified rejuvenation, which is defined as a set of short transdimensional rjMCMC chains. This means that any dimension adjustment is only possible during rejuvenation.
Note that with this modification, rejuvenation tackles the task of transdimensional convergence and the task of avoiding particle impoverishment the same time. This could result in an increased level of filter degeneracy and to a non-converging particle filter. To overcome this problem as one solution is the use of tempered probability distribution for resampling, for a more gradual convergence over the timesteps \cite{Beskos2014}. A different way of handling this problem could be the separation of these two tasks, by introducing an additional in-dimension rejuvenation step to avoid impoverishment.\\
It is also important to note, that because the rjMCMC chains run on the individual particles, the ensemble very likely have members of different dimensions at the same time. This introduces some practical complications, for example the formulation eq.\ref{particle_density} is not valid anymore in a strict sense. The statistical analysis of such datasets could potentially require data conversion techniques \cite{Brooks2003,Somogyvari2019}.\\
Note that this extension of the methodology is suitable to augment any type of particle filter which uses stationary Markov kernels for rejuvenation (e.g. the hybrid Ensemble Kalman-filter).

\section{Model examples}
In this section we present two case studies for the proposed transdimensional data assimilation methodology. Both examples are strongly simplified, as they serve mainly development and demonstration purposes.

\subsection{Fourier analysis}
Consider the following example. We are approaching to an area where a set of radio stations are transmitting signals continuously on specific frequencies. We have a receiver which observes the radio signal in the temporal domain. Initially only a few of the stations can be observed, but as we are getting closer, their numbers are increasing. Also because of the approach the observed signal amplitude is increasing. Our task is to decompose this signal and identify the individual signal sources.

This example is a one-dimensional signal analysis based on Fourier transformation. A dynamically changing signal is defined in the frequency space as the state model which is observed continuously in the time domain. The different frequencies of the signal continuously increase their amplitudes and at random timesteps new frequency modes appear in the signal. This gives a simple transdimensional dynamicity to the system. The objective of the particle filter is to reproduce the signal, and to forecast the state space model in the frequency domain.
The state space vector (parameter model) is defined by finite number ($L$) of Fourier modes ($m^l$), that represent the frequencies of the signal:

\begin{equation}
	\{m_t^l\} = m_t^0,m_t^1...m_t^{L-1}
\end{equation}

The number of modes is increasing over time. In the shown example initially three Fourier modes exist and every 10th timestep a new mode is added to the model. This addition is only performed until the maximum allowed mode number is reached, which is 10 in the example.
The initial values of $m_0^l$ are randomly drawn from a predefined range with uniform distribution. Their values are not constant, it changes over time according to a gain vector:

\begin{equation}
	\{v_t^l\} = v_t^0,v_t^1...v_t^{L-1}
\end{equation}

The gain values are given when a new mode is created, drawn from a random distribution. The state space model preserves these values over time. In the example case only positive values are used, so the amplitude of the signal is monotonously increasing over time. The observation model is defined as the signal calculated from the Fourier modes using the inverse discrete Fourier transformation:

  \begin{equation}
        \xi_{F,t} (x) = \frac{1}{L} \sum_{l=0}^{L-1}{m^l exp\left(\frac{i2\pi}{L}lx\right)}
  \end{equation}

where $\xi_{F,t}(x)$ is the observed signal at time $t$. To avoid any confusions with the temporal dynamics, we talk about the signals in the spatial domain $x$ (instead of time) hence the used Fourier modes represent spatial frequencies. To make the example realistic, random Gaussian noise is added to the observed signal.
Figure 1 shows the used example model. As time progresses the observed signal becomes more complex because of the newly introduced frequency modes. Different amplitude gains on different frequencies distort the signal shape as their relative amplitudes changing. The goal of the particle filter is to capture this evolution and recreate the signal as well as possible.

\begin{figure}[htp]
\begin{center}
  \includegraphics[width=5in]{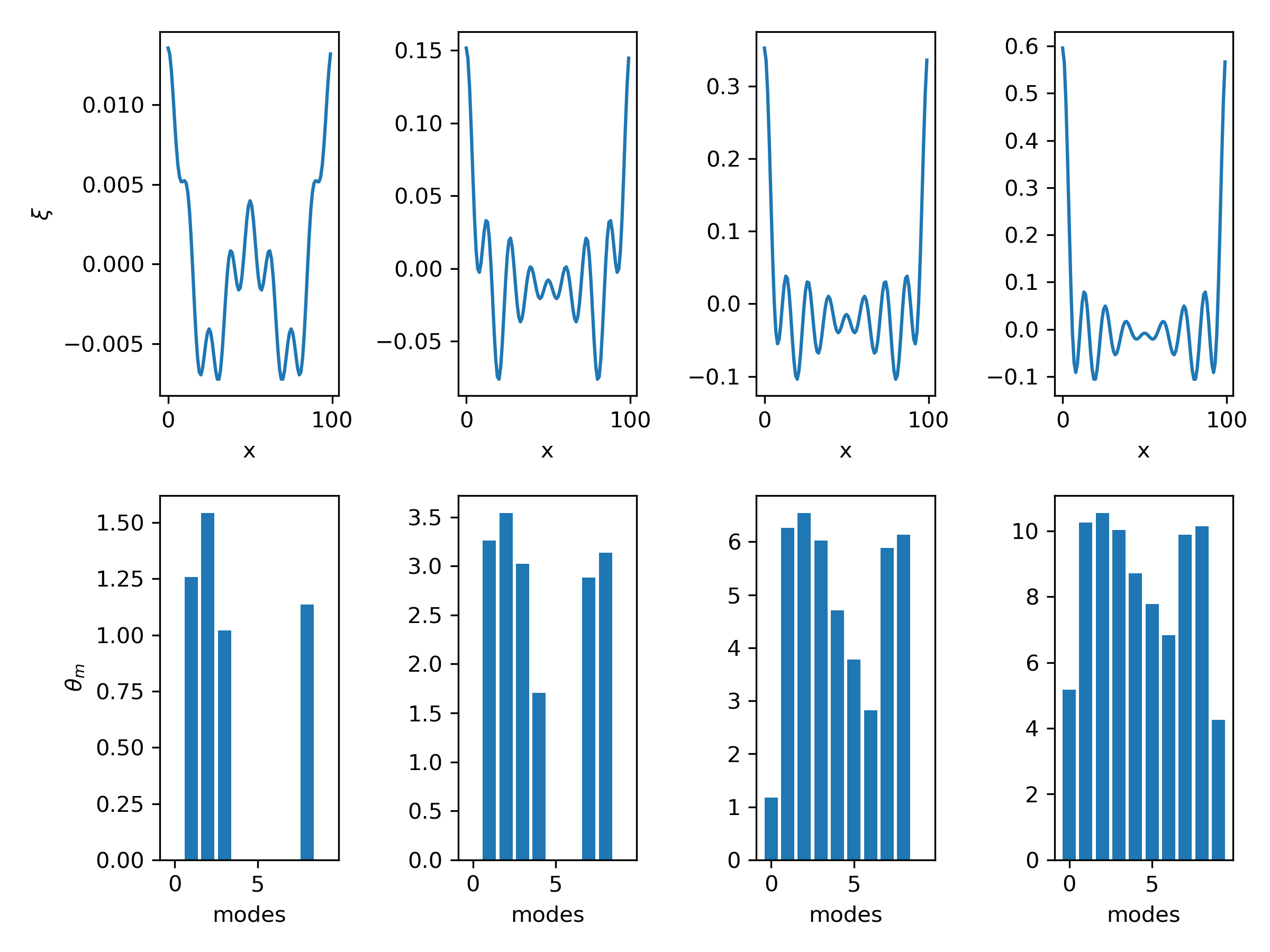}\\
  \caption{Evolution of the exemplary frequency model over time. The upper row shows the observed signal, the lower row shows the underlying state space model at timesteps t=0,25,50 and 75.}\label{F_model}
  \end{center}
\end{figure}

The transdimensional particle filter uses the observed signals in Figure \ref{F_model} as data input and tries to infer the frequency modes. At every timestep the filter forecasts the state of the state space model. These forecasts are evaluated by comparing to the observed signal in the particle reweighting and resampling step. After these steps the particles could get rejuvenated with the transdimensional rjMCMC. In this implementation rejuvenation is only at every 5th timestep (starting from step 3) with 10 rjMCMC updates on each particle. At each rjMCMC update, mode addition, mode deletion or mode value change could happen, with equal probabilities. The applied modification is then evaluated by the Metropolis Hastings Green acceptance criteria \cite{Green1995}.

\begin{figure}[htp]
\begin{center}
  \includegraphics[width=5in]{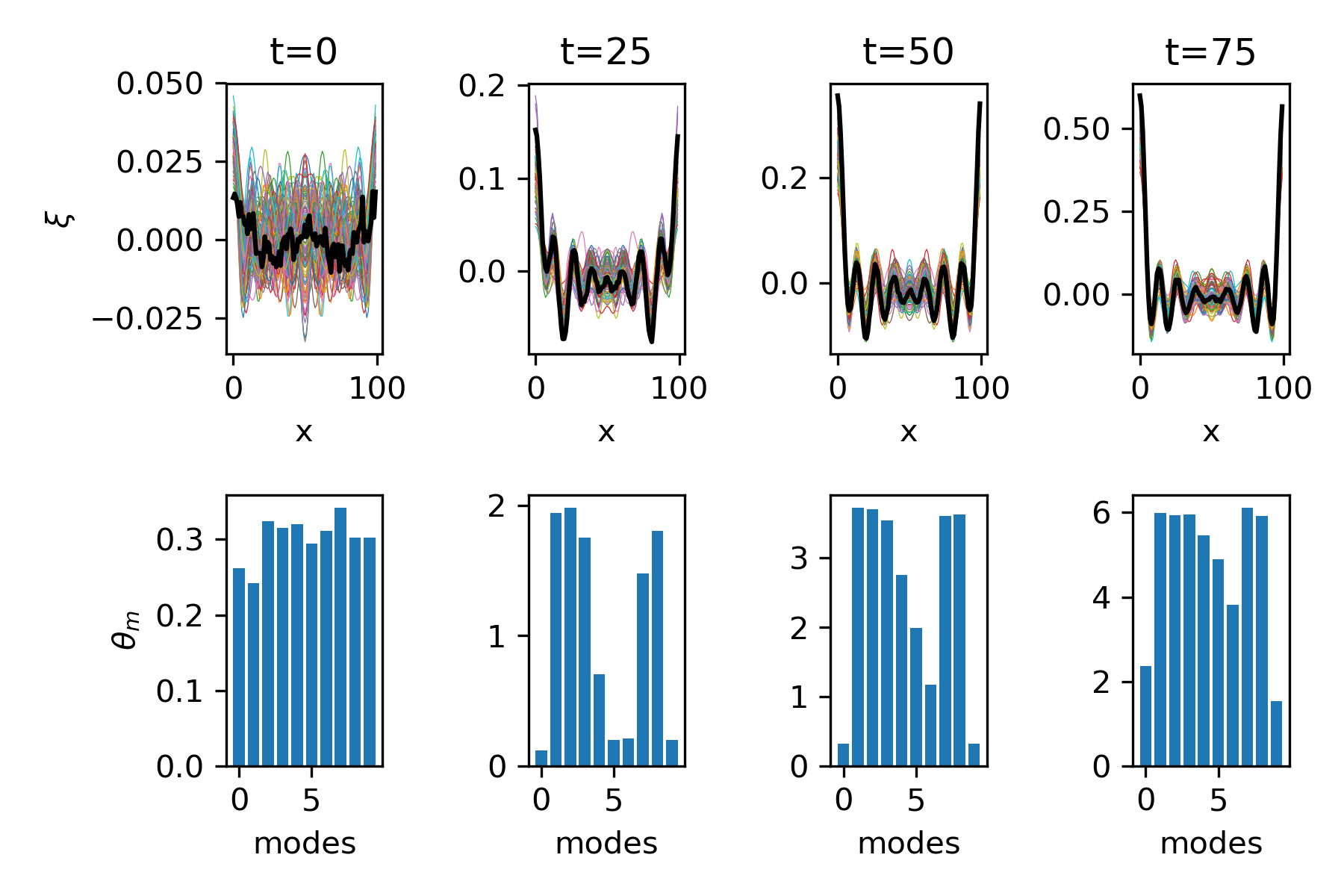}\\
  \caption{Reconstructions of the Fourier signals. The upper row shows the reconstructed signals with colored lines and the observed reference signal in black on top. The lower row shows the mean of the reconstructed state space vectors. The depicted timesteps are t=0,25,50 and 75.}\label{F_result}
  \end{center}
\end{figure}

100 particles were used for the inference, each defined by a state space vector ${m_t^n }$ and by a gain vector ${v^n }$. The target probability distribution for reweighting and rejuvenation was defined as:

\begin{equation}
	P(m_t^n|\xi_{obs}(x)) = \frac{1}{\sqrt{2 \pi \sigma^2}}exp \left( -\frac{1}{2}\frac{(\xi_{sim}(x)-\xi_{obs}(x))^2}{\sigma^2}\right)
\end{equation}

where we use the simplified notation $\xi_{obs}(x)$ for the observed signals, and  $\xi_{sim}(x)$ from the simulated signals over the state space vector.
The acceptance rate of the transdimensional Markov kernel therefore calculated as:
\begin{equation}
\alpha(x,y)= \frac{P(m_t^{*,n}|\xi_{obs}(x))q(m_t^n|m_t^{*,n})}{P(m_t^{n}|\xi_{obs}(x))q(m_t^{*,n}|m_t^n)} 1
\end{equation}

where $m_t^{*,n}$ is the proposed new model realization of an MCMC update. This kernel makes the MCMC reversible, and therefore the rejuvenation stationary. Note that the Jacobian equals to one, this is because the possible transdimensional updates are all discrete \cite{Denison2002}.\\
Because this example showed strong particle impoverishment otherwise, different probability distributions were used within resampling and rejuvenation ($\sigma_{res} = 0.03$, $\sigma_{rej} = 0.001$).
Because of the relatively light computational burden assimilating 100 timesteps with the filter only took a few minutes on a laptop without any parallelization. The results of the assimilation are shown in Fig. \ref{F_result}.

Figure \ref{F_result} shows results from four different timesteps from the reconstruction by the particle filter. The upper row shows the observed signal as a black line and the reconstructed signals of the different particles as colored lines behind. Note the noise of the observed signal at the initial timesteps and see as it disappears with the increase of amplitudes. 
The first column depicts the 0th timestep, where no data had been assimilated yet. The signals of the different particles are spread out without any connection to the observation. This spread reduces over time as the different particles are getting closer to each other and to the observation. They do not overlap completely, showing that the implemented rejuvenation obstructs the impoverishment of the filter.
Compared to the reference, the relation between the reconstructed mean amplitudes is very similar, but their absolute values differ. This is an artifact of the averaging over the ensemble and the fact that the signal is more sensitive to the relative mode amplitudes. By choosing a different metric for calculating the misfit the algorithm could be made more sensitive to the absolute amplitudes.

\begin{figure}[htp]
\begin{center}
  \includegraphics[width=4in]{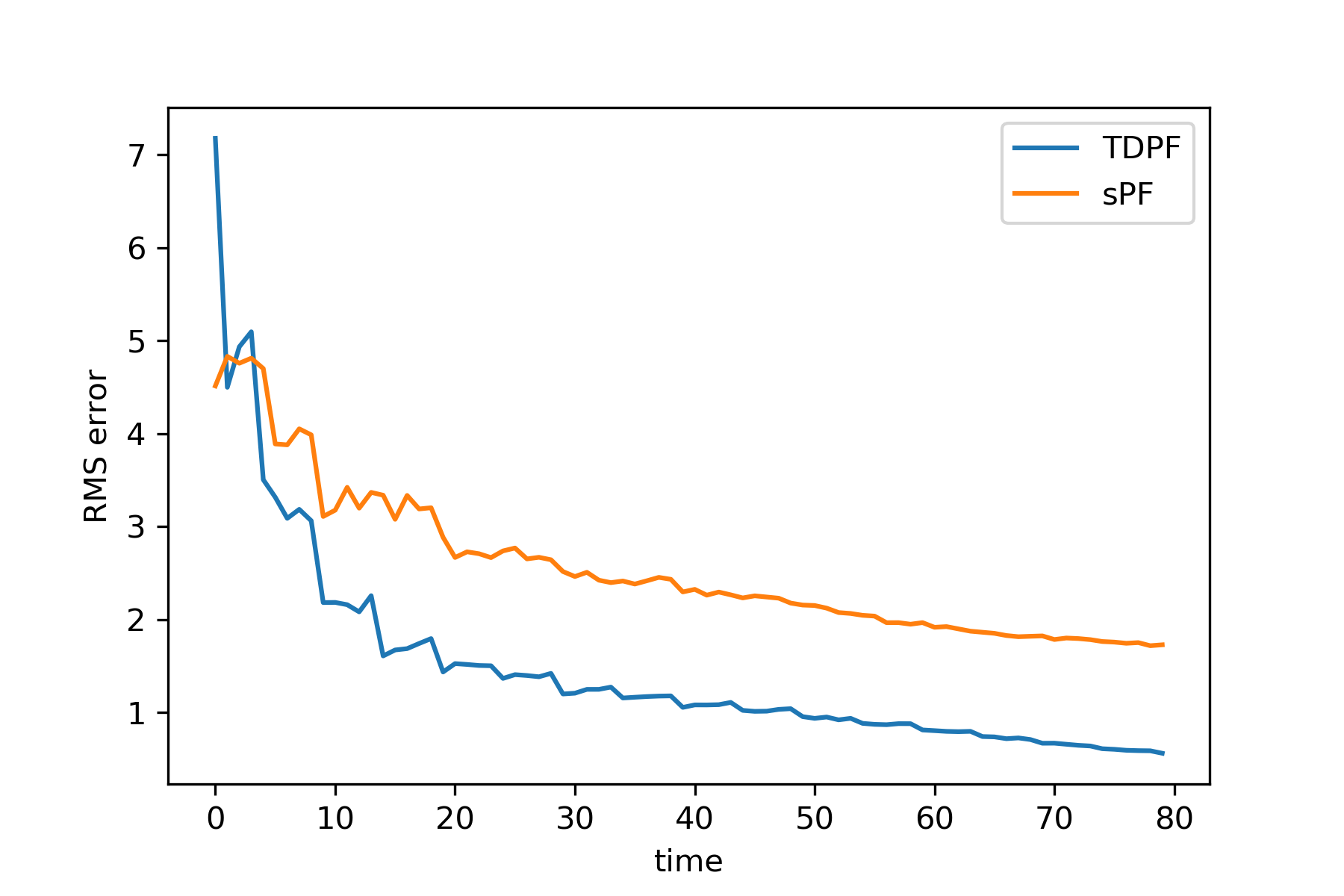}\\
  \caption{Convergence of the transdimensional particle filter (TDPF) versus the standard bootstrap filter (sPF).}\label{F_compare}
  \end{center}
\end{figure}

To demonstrate the performance of the filter Fig. \ref{F_compare} shows the evolution of the mean RMSE misfit of the observations of the transdimensional particle filter compared to the standard particle filter.

The only difference between the two shown filters is that the transdimensional updates are disabled in the rejuvenation of the standard particle filter. All the other parametrization remained the same. In the very early timesteps the standard particle filter often outperforms the transdimensional version. Here, if the initial realizations are close to the target dimensionality, the transdimensional filter makes unnecessary dimensional jumps which gets rejected. The standard particle filter could use these updates for convergence to the target distribution. At later timesteps, where the dimensionality changes, the transdimensional filter consequently outperforms the standard one.

The convergence plot of the filter can also be used to check the different parametrization of the particle filter and the MCMC chains. If the particle filter performs poorly, there is no decreasing trend visible only periodic drops where the MCMC rejuvenation corrects the algorithm. In Fig. \ref{F_compare} there is no strong sign of this periodicity, showing that the two parts of the method are in agreement, and that the data assimilation results are improving over time.

Running the data assimilation with different setups showed:
\begin{itemize}
	\item	Reducing the number of the rejuvenation phases result in particle degeneration at later timesteps. After this happens, the particle system cannot react to any major changes within the model, and large variations start to appear in the As the original aim of rejuvenation is to mitigate particle impoverishment, it is obvious that more of these phases remove its effects. Running the algorithm with more particles could also help, but degeneration would still return at larger timesteps. This happens with the standard particle filter ass well, however less frequently.
	\item	Adding noise improve the stability of the algorithm. Removing it completely led to incorrect assimilation and a non-converging system.
	\item	Choosing too large variance values for reweighting and rejuvenation lead to non-converging results. Choosing too small values lead to particle impoverishment.
	The results imply that this simple model can be captured well, compared to the models in Fig. \ref{F_model}. Still the example not necessarily shows the full potential of the transdimensional extension, as the signal can relatively well reconstructed only by using a few Fourier modes and ignoring most of them. This happens with the standard particle filter in Fig. \ref{F_compare} which still shows convergence despite its fixed dimensionality.
\end{itemize}

\subsection{Object tracking}
Object tracking problems are classic subjects of data assimilation algorithms (see e.g. (Gordon et al., 1993a \cite{Gordon1993a})). In these applications the task of data assimilation is to infer the coordinates and velocity vectors of moving objects from sparse observations. 

In this synthetic example we use a simulated radar image to track a set of flying objects (e.g. airplanes). These objects could randomly appear and disappear on the radar, meaning that the state space vector size could change over time. This is considered by the transdimensional extension of the particle filter.

\begin{figure}[htp]
\begin{center}
 \includegraphics[width=4in]{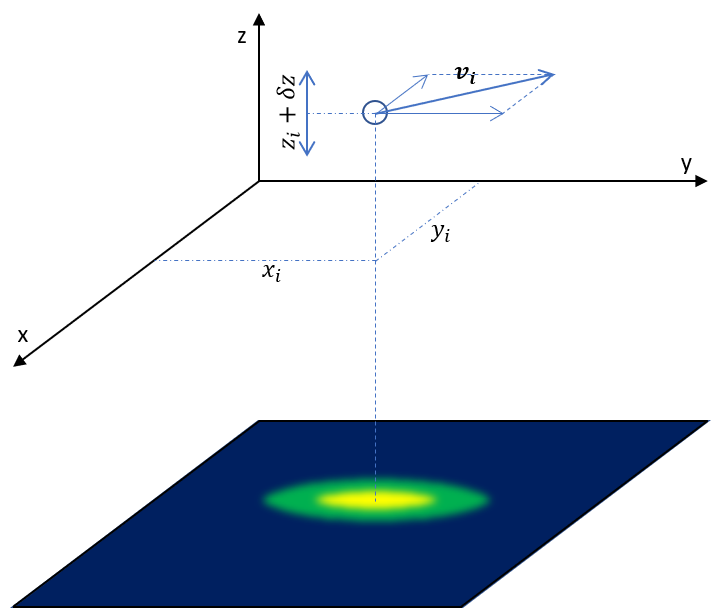}\\
  \caption{Concept of the object tracking model example (flying planes within a 3-D domain - top), and its observation model (2-D radar image  - bottom).}\label{OT_model}
\end{center}
\end{figure}

In this simple example, the objects are moving according to a predefined velocity vector with a randomly chosen direction and length. Hence the state vector for each object reads as:
\begin{equation}
\boldsymbol{\Theta}_t^i = [x_t^i,y_t^i,z_t^i,\boldsymbol{v}_t^i]
\end{equation}

If an object descends below a critical $z$ value, it gets deleted from the state space, modelling that it has landed ($z_{crit}=0.1$, a detection limit for the radar). Because the velocity vector is defined in 2-D, changes in object elevations are randomly drawn: $\delta z \sim N(0,|\boldsymbol{v}_t^i |)$.

For the simulaton at $t=0$ a random number of objects are initialized (maximum 10). Every timestep, a new object could appear at a random location in the domain, with $20\%$ probability. The velocity vector of new objects is randomly drawn, the amplitude from $|\boldsymbol{v}_t^i | \sim N(0.1,1)$ and direction $\gamma_v \sim N(1,1)$. The velocity values change randomly at every timesteps, but the directions are kept permanent. The simulation was run for 50 timesteps and its results are shown in Figure 5.

\begin{figure}[htp]
\begin{center}
 \includegraphics[width=5in]{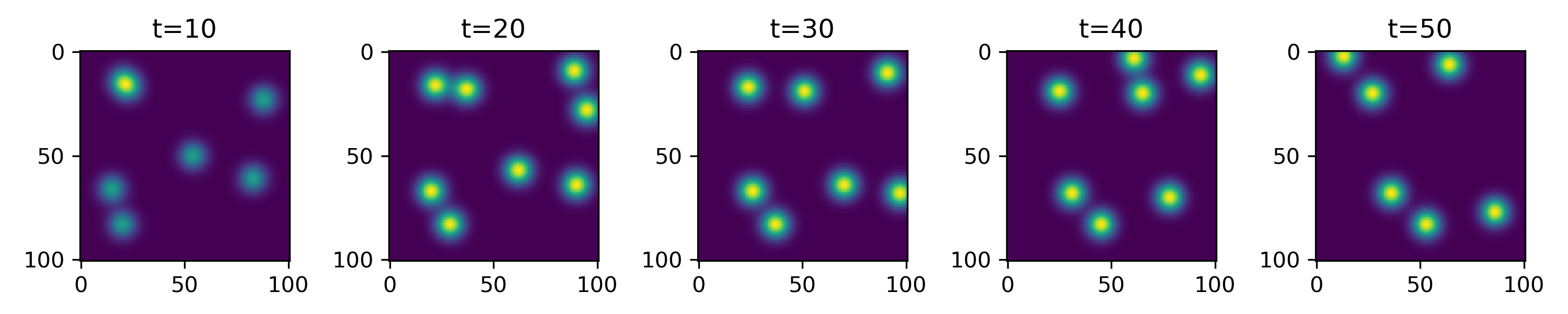}\\
  \caption{Snapshots of radar images from the object tracking example.}\label{radar_example}
\end{center}
\end{figure}

Figure 5 shows snapshots of the observed radar images at different timesteps. Each observation is a rasterized version of the object locations, without considering the vertical component. To simulate a radar image ($\xi_R$), the rasterization of the objects is done by the superposition of 2-D Gaussian functions:

\begin{equation}
\xi_r(x,y) = \sum_{i=1}^N \frac{1}{\sqrt{2\pi\sigma_i^2}}exp\left[ - \frac{1}{2} \frac{(x_i-x)^2+(y_i-y)^2}{\sigma_i^2}\right]
\end{equation}

In the example, the resolution of the radar grid is 100x100 pixels, and the variance of the used gaussian functions are 20 pixels. The input of the particle filter are the radar image matrices over the 50 timesteps.
The radar images do not contain any information about the vertical coordinate of the tracked objects (which is a realistic assumption for a simple radar system), so the state space vector in the particle filter can be set as:

\begin{equation}
\boldsymbol{\theta}_t^i = [x_t^i,y_t^i,\boldsymbol{v}_t^i]
\end{equation}

The state space vector consist of all the objects together, and every change in its dimensionality means the addition or the deletion of 3 state variables. For the data assimilation the particle filter is set up with 100 particles. Rejuvenation was used at every 10th timestep, starting from step 5 (5, 15, 25…). The rejuvenation used 30 rjMCMC iterations on each of the particles. Compared to the previous example this is a larger number, which is needed because of the increase in state variables (4 parameter per dimension compared to the previous 1). For reweighting and rejuvenation, a similar Gaussian target distribution function was used as in the previous example (for timestep $\tau$):

\begin{equation}
P(\theta_t^i|\xi_R(t)) = \frac{1}{\sqrt{2\pi\sigma^2}}exp\left[ - \frac{1}{2} \frac{(\xi(\theta_t^i)-\xi_R(t))^2}{\sigma^2}\right]
\end{equation}

and the acceptance rate of the transdimensional Markov kernel is also similar:
\begin{equation}
\alpha(x,y)= \frac{P(\theta_t^{*,i}|\xi_{R}(t))q(\theta_t^i|\theta_t^{*,i})}{P(\theta_t^{i}|\xi_{R}(t))q(\theta_t^{*,i}|\theta_t^i)} 1
\end{equation}

 Similarly to the previous example, an efficient model run required the tempering of the target distribution at the reweighting step of the reweighting compared to the rejuvenation. Our experience showed that choosing a magnitude lower variance values for the reweighting results in an optimal convergence rate without the risk of particle impoverishment.
 
 \begin{figure}[htp]
\begin{center}
 \includegraphics[width=5in]{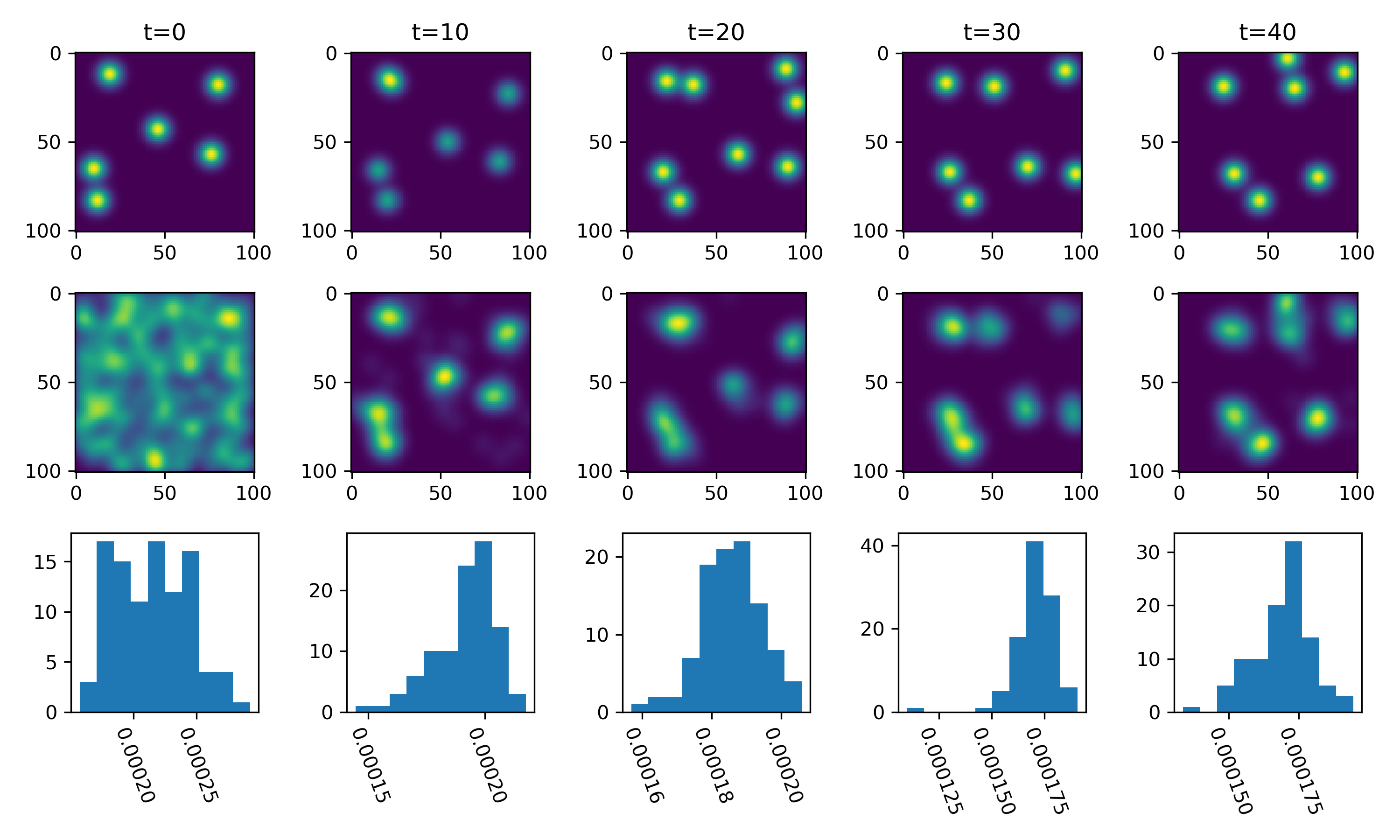}\\
  \caption{Data assimilation results of the object tracking example. TOP – radar observations, MIDDLE – reconstructed object locations, BOTTOM – distribution of errors within the particle set.}\label{radar_result}
\end{center}
\end{figure}

The results of the data assimilation are shown in figure 6. In this example, the state space cannot be visualized as directly as previously because of the differences in dimensions. To have a comparable representation, we plot the mean of the observations over the complete particle set as a probability map. This is obtained by generating the individual radar images and stacking them together (similar approach was used by (Bodin and Sambridge (2009) \cite{Bodin2009}; Jiménez et al., (2016) \cite{Jimenez2016}; Somogyvári et al., (2017) \cite{Somogyvari2017}). The probability maps can be directly compared to the original observed radar images. This visualization very well represents any spatial uncertainties within the particle set as lower probability objects appear with lower intensity. Additionally we plot the distribution of the RMS error within the particle set. This could give an indication if the particles set is collapsing into a single particle.

The 0th timestep shows the unconstrained initial ensemble. At this point we can see the initial locations of the objects of the particles, that show a close to uniform distribution over the domain. This is also clear from the distribution of the particle misfits, which is shown here in a histogram. At the 10th timestep the particles have been already resampled 9 times, and they are also over the first rejuvenation with the rjMCMC chains. At this point the reconstruction is very accurate, and the error distribution is very narrow. This is changed again at the 20th timestep, where new objects appear again. In the top left corner two objects are very close to each other which in the reconstruction appear as one, but with an increased spatial uncertainty. The later timesteps are similar, generally most of the objects are located, but freshly appeared or disappeared ones cause uncertainties. This results in a relatively shaky convergence curve, in which the average misfits at each timestep are plotted (Fig.7). The plot also shows the convergence of the individual particles, showing that the overall trends are very similar to each other.

 \begin{figure}[htp]
\begin{center}
 \includegraphics[width=4in]{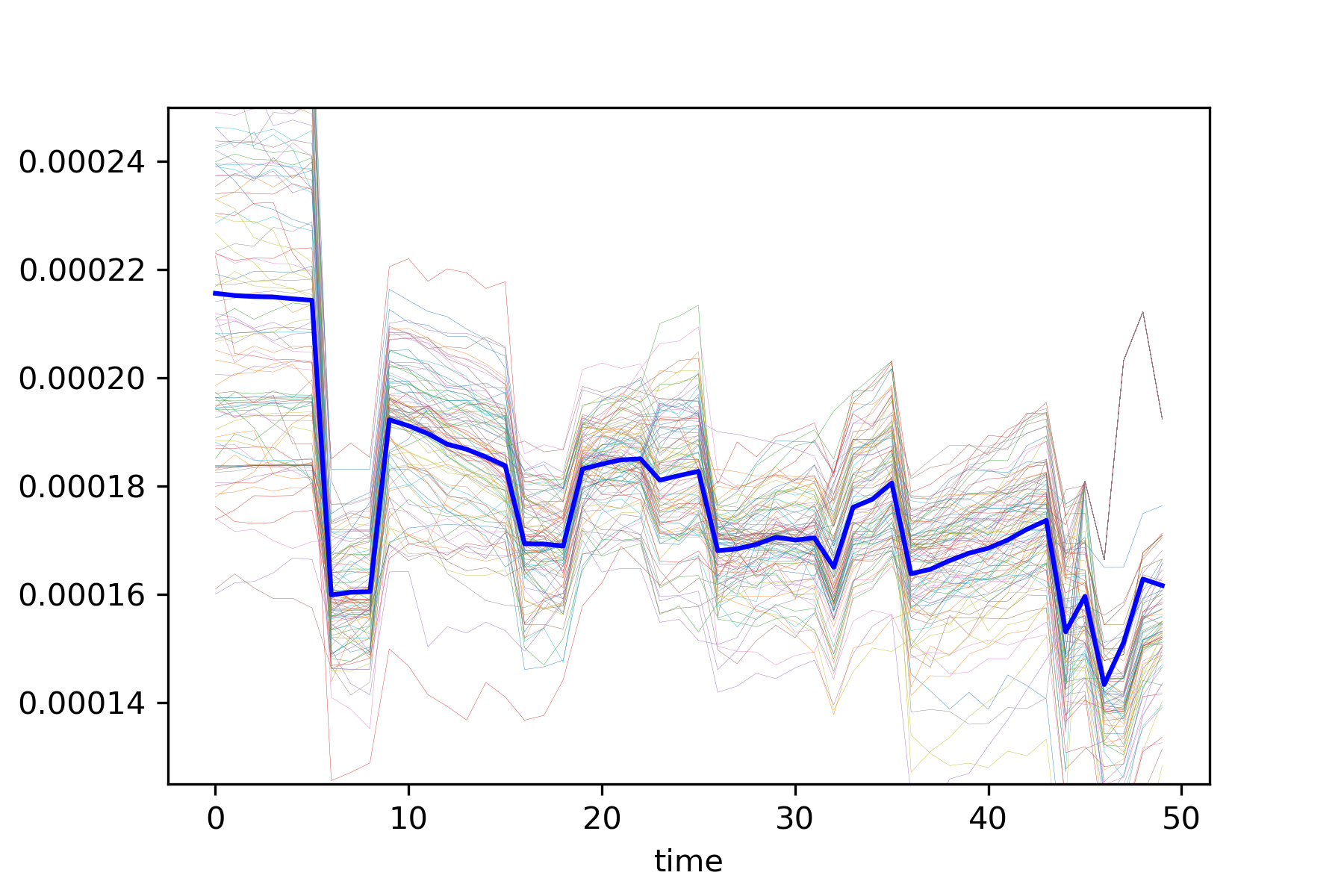}\\
  \caption{Convergence of the particle filter.}\label{convergence}
\end{center}
\end{figure}

The rjMCMC rejuvenation happen at timesteps 5,15,25,35 and 45. Drops of the convergence are visible at these timesteps. The drops however are not exclusive to the rejuvenation, showing the performance of the reweighting-resampling part of the particle filter as well. Our experience shown, that the exact shape of this curve is strongly dependent on the observed problem and less of a particular particle filter run, proving that the algorithm is robust. The appearance or disappearance of many objects in a short period of time, could temporarily increase the misfit, which then can only be corrected in the next rejuvenation phase. One possible solution to this problem could be to use misfit-triggered rejuvenation, activating this phase of the algorithm when a significant increase in the misfit is observed. Existing particle filter implementations often use a similar strategy to cope with filter impoverishment \cite{Andrieu2010}.

 \begin{figure}[htp]
\begin{center}
 \includegraphics[width=5in]{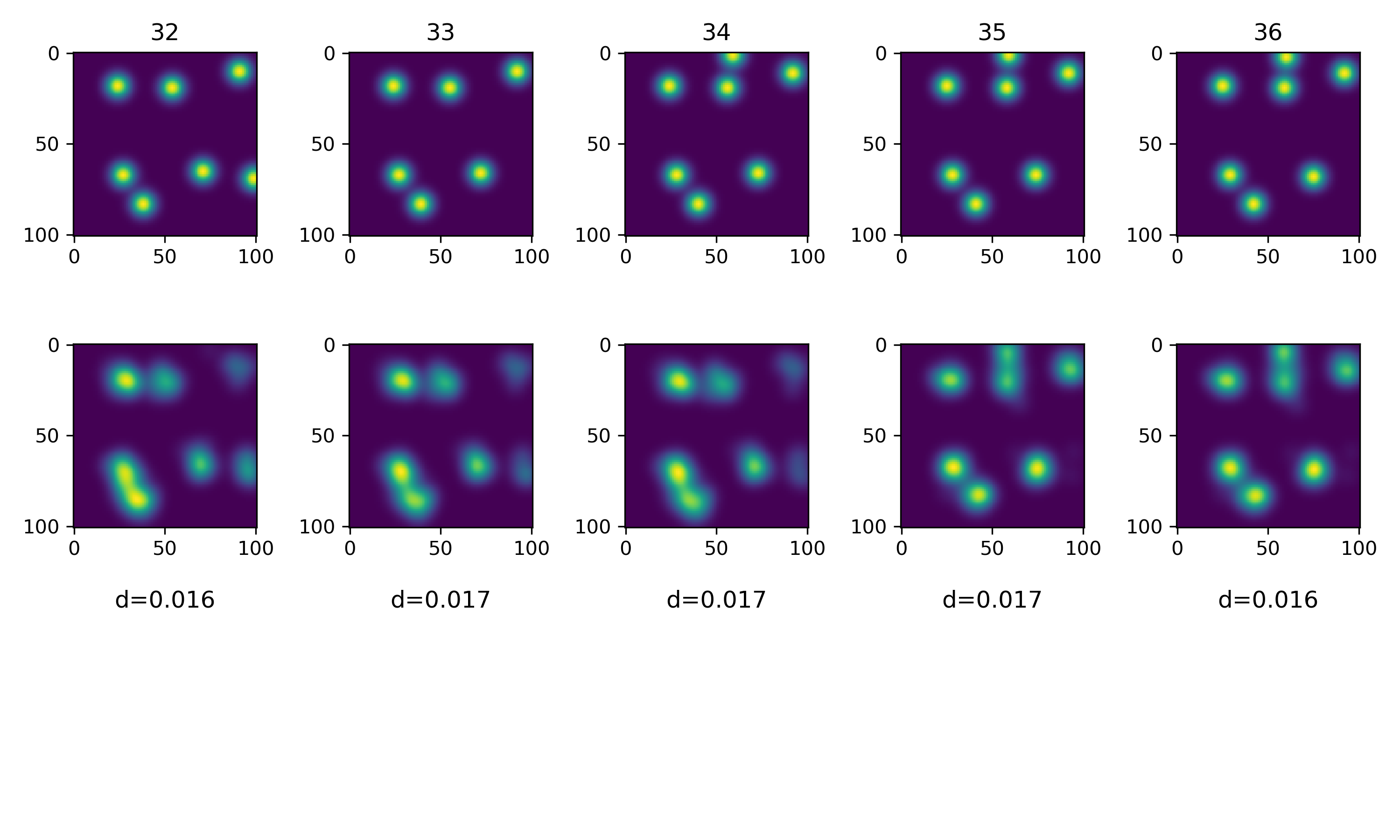}\\
  \caption{Detailed analysis of data assimilation timesteps. TOP: reference model, BOTTOM: reconstruction.}\label{detailed_steps}
\end{center}
\end{figure}

Figure 8 shows some of the individual timesteps. At step 33, the bottom right object disappears from the radar. This results in an increase in the misfit value. The system only recovers after step 35, where the transdimensional MCMC removes the extra object from the reconstruction (note that the shown misfit values are calculated after resampling, before the rejuvenation).

\section{Conclusions}
In this study we have demonstrated the potential of transdimensional data assimilation by combining particle filters with the reversible-jump Markov chain Monte Carlo algorithm. The presented technique is capable of changing the dimensionality of the state space vector on the fly, giving extra flexibility compared to existing data assimilation methods.

One additional advantage of the transdimensional particle filter that we have found is, that the models of the individual particles could run on lower dimensions than it is required to describe the full problem. In our object tracking example, the individual model realizations (the particles) always consist of only 2-3 objects, even when the observed system has more than five (see Figure 9). This results in a significant increase in computational efficiency and inversion robustness, without compromising too much of the description of the investigated system. 

 \begin{figure}[htp]
\begin{center}
 \includegraphics[width=5in]{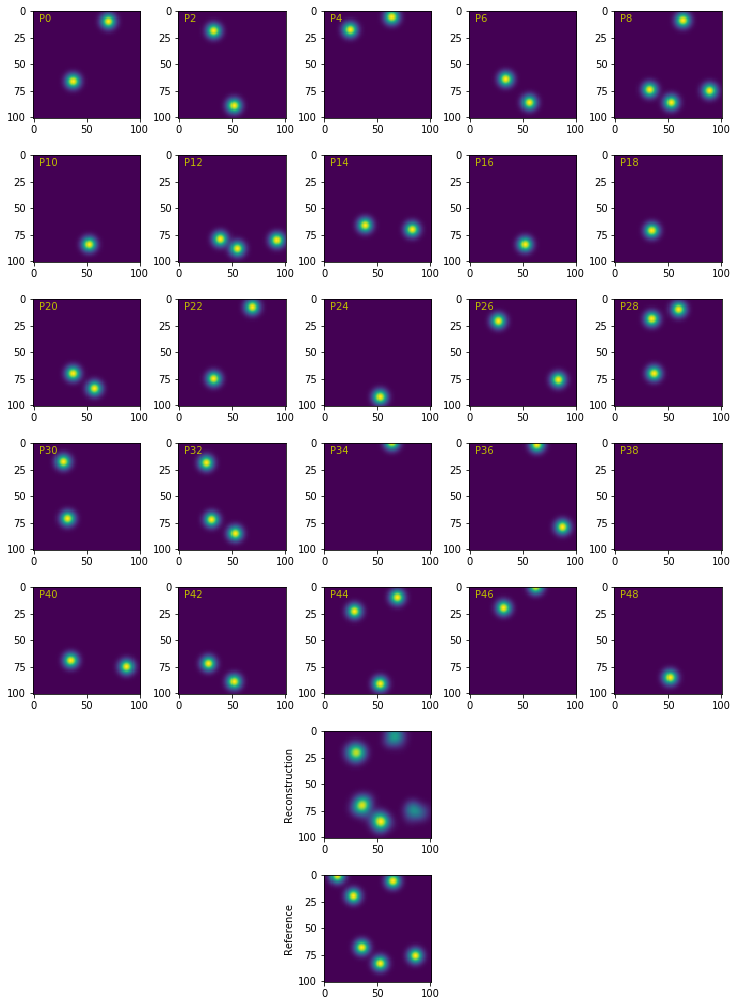}\\
  \caption{Individual realizations within the particle set of the object tracking example.}\label{particles}
\end{center}
\end{figure}

One potential application field of the presented methodology are physical problems with adaptive discretization. In the context of real-world spatial problems, adaptive discretization could allow for refined model grids at sensitive parts of the model, and course grids at less relevant parts. By adapting the discretization to the problem, better numerical results could be obtained without compromising computational times.

In data assimilation problems, the main part of computational time is running the forward models, therefore these improvements would be a great step in effectiveness. The reduction in the number of model parameters could also reduce the freedom of the inverse problem \cite{Jimenez2016}.

Choosing the particle filter for transdimensional data assimilation in this study, was not motivated by the classic non-linearity argument of the particle filter, but by the simple integration via the particle rejuvenation method. Henceforth, it would be interesting to explore the integration to other data assimilation methods, such as the EnKF. EnKF is one of the most widely used data assimilation technique, with better performance due to its more controlled inversion scheme but suffers from a similar filter degeneracy problem in higher dimensions. There exist methodologies that combines particle filters with EnKF to improve filter efficiency \cite{VanLeeuwen2019}. These hybrid approaches could be extended for transdimensional problems the same way as it is shown in this paper.

\section*{Acknowledgments} 
This publication was financially supported by Geo.X, the Research Network for Geosciences in Berlin and Potsdam under the Grant Number SO\_ 087\_ GeoX. This research has been also supported by the Deutsche Forschungsgemeinschaft (DFG) under Grant CRC 1294 “Data Assimilation (Project B04)”. Special thanks to Paul Rozdeba and Theresa Lange for the support with the preparation of this manuscript.

\printbibliography

@article{Li2013,
author = {Li, T and Sun, S and Sattar, TP and Corchado, JM},
journal = {Expert Systems with Applications},
pages = {1--34},
title = {{Fighting against Sample Degeneracy and Impoverishment in Particle Filters: Particularly on Intelligent Choices}},
url = {http://arxiv.org/abs/1308.2443},
year = {2013}
}

@article{Morzfeld2017,
archivePrefix = {arXiv},
arxivId = {1512.03720},
author = {Morzfeld, Matthias and Hodyss, Daniel and Snyder, Chris},
doi = {10.1080/16000870.2017.1283809},
eprint = {1512.03720},
issn = {16000870},
journal = {Tellus, Series A: Dynamic Meteorology and Oceanography},
number = {1},
pages = {1--15},
publisher = {Taylor {\&} Francis},
title = {{What the collapse of the ensemble Kalman filter tells us about particle filters}},
url = {http://dx.doi.org/10.1080/16000870.2017.1283809},
volume = {69},
year = {2017}
}

@article{Somogyvari2017,
author = {Somogyv{\'{a}}ri, M{\'{a}}rk and Jalali, Mohammadreza and {Jimenez Parras}, Santos and Bayer, Peter},
doi = {10.1002/2016WR020293},
issn = {00431397},
journal = {Water Resources Research},
month = {jun},
number = {6},
pages = {5104--5123},
title = {{Synthetic fracture network characterization with transdimensional inversion}},
url = {http://doi.wiley.com/10.1002/2016WR020293},
volume = {53},
year = {2017}
}

@article{Denison2002,
author = {Denison, D. G.T. and Adams, N. M. and Holmes, C. C. and Hand, D. J.},
doi = {10.1016/S0167-9473(01)00073-1},
issn = {01679473},
journal = {Computational Statistics and Data Analysis},
number = {4},
pages = {475--485},
title = {{Bayesian partition modelling}},
volume = {38},
year = {2002}
}

@article{Kalman1961,
author = {Kalman, R. E. and Bucy, R. S.},
doi = {10.1115/1.3658902},
issn = {1528901X},
journal = {Journal of Fluids Engineering, Transactions of the ASME},
number = {1},
pages = {95--108},
title = {{New results in linear filtering and prediction theory}},
volume = {83},
year = {1961}
}

@article{Bodin2009,
author = {Bodin, Thomas and Sambridge, Malcolm},
doi = {10.1111/j.1365-246X.2009.04226.x},
isbn = {0956540X$\backslash$r1365246X},
issn = {0956540X},
journal = {Geophysical Journal International},
number = {3},
pages = {1411--1436},
title = {{Seismic tomography with the reversible jump algorithm}},
volume = {178},
year = {2009}
}

@article{Jimenez2016,
author = {Jim{\'{e}}nez, S and Mariethoz, G and Brauchler, R and Bayer, P},
doi = {10.1002/2015WR017922},
issn = {00431397},
journal = {Water Resources Research},
month = {may},
number = {5},
pages = {3966--3983},
title = {{Smart pilot points using reversible-jump Markov-chain Monte Carlo}},
url = {http://doi.wiley.com/10.1002/2015WR017922},
volume = {52},
year = {2016}
}

@incollection{Geyer2011,
author = {Geyer, Charles},
booktitle = {Handbook of Markov Chain Monte Carlo},
chapter = {1},
doi = {10.1201/b10905},
editor = {Brooks, Steve and Gelman, Andrew and Jones, Galin and Meng, Xiao-Li},
isbn = {978-1-4200-7941-8},
number = {3},
pages = {45},
publisher = {Chapman and Hall/CRC},
title = {{Introduction to Markov Chain Monte Carlo}},
volume = {20116022},
year = {2011}
}

@article{Sambridge2006,
author = {Sambridge, Malcolm and Gallagher, K. and Jackson, A. and Rickwood, P.},
doi = {10.1111/j.1365-246X.2006.03155.x},
isbn = {0956540x},
issn = {0956540X},
journal = {Geophysical Journal International},
number = {2},
pages = {528--542},
title = {{Trans-dimensional inverse problems, model comparison and the evidence}},
volume = {167},
year = {2006}
}

@article{Geweke1989,
author = {Geweke, John},
doi = {10.2307/1913710},
issn = {00129682},
journal = {Econometrica},
month = {nov},
number = {6},
pages = {1317},
title = {{Bayesian Inference in Econometric Models Using Monte Carlo Integration}},
url = {https://www.jstor.org/stable/1913710?origin=crossref},
volume = {57},
year = {1989}
}

@article{Fearnhead2017,
archivePrefix = {arXiv},
arxivId = {1709.04196},
author = {Fearnhead, Paul and K{\"{u}}nsch, Hans},
doi = {10.1146/annurev-statistics-031017-100232},
eprint = {1709.04196},
month = {sep},
pages = {1--31},
title = {{Particle Filters and Data Assimilation}},
url = {http://arxiv.org/abs/1709.04196{\%}0Ahttp://dx.doi.org/10.1146/annurev-statistics-031017-100232 http://arxiv.org/abs/1709.04196 http://dx.doi.org/10.1146/annurev-statistics-031017-100232},
year = {2017}
}

@article{Sambridge2012,
author = {Sambridge, M. and Bodin, T. and Gallagher, K. and Tkalcic, H.},
doi = {10.1098/rsta.2011.0547},
isbn = {1364503X (ISSN)},
issn = {1364-503X},
journal = {Philosophical Transactions of the Royal Society A: Mathematical, Physical and Engineering Sciences},
month = {dec},
number = {1984},
pages = {20110547--20110547},
pmid = {23277604},
title = {{Transdimensional inference in the geosciences}},
url = {http://rsta.royalsocietypublishing.org/cgi/doi/10.1098/rsta.2011.0547},
volume = {371},
year = {2012}
}

@misc{Chopin2002,
author = {Chopin, Nicolas},
booktitle = {Biometrika},
doi = {10.1093/biomet/89.3.539},
issn = {00063444},
number = {3},
pages = {539--551},
pmid = {178151800004},
title = {{A sequential particle filter method for static models}},
volume = {89},
year = {2002}
}

@article{Andrieu2010,
archivePrefix = {arXiv},
arxivId = {arXiv:1208.5721},
author = {Andrieu, Christophe and Doucet, Arnaud and Holenstein, Roman},
doi = {10.1111/j.1467-9868.2009.00736.x},
eprint = {arXiv:1208.5721},
isbn = {1467-9868},
issn = {13697412},
journal = {Journal of the Royal Statistical Society Series B-Statistical Methodology},
number = {3},
pages = {269--342},
pmid = {18825524},
title = {{Particle Markov chain Monte Carlo methods}},
url = {http://links.isiglobalnet2.com/gateway/Gateway.cgi?GWVersion=2{\&}SrcAuth=mekentosj{\&}SrcApp=Papers{\&}DestLinkType=FullRecord{\&}DestApp=WOS{\&}KeyUT=000277976300001},
volume = {72},
year = {2010}
}

@article{Brooks2003,
archivePrefix = {arXiv},
arxivId = {arXiv:1011.1669v3},
author = {Brooks, S.P and Giudici, P and Philippe, A},
doi = {10.1198/1061860031347},
eprint = {arXiv:1011.1669v3},
isbn = {9788578110796},
issn = {1061-8600},
journal = {Journal of Computational and Graphical Statistics},
month = {mar},
number = {1},
pages = {1--22},
pmid = {25246403},
title = {{Nonparametric Convergence Assessment for MCMC Model Selection}},
url = {http://www.tandfonline.com/doi/abs/10.1198/1061860031347},
volume = {12},
year = {2003}
}

@incollection{Doucet2001a,
address = {New York, NY},
author = {Doucet, Arnaud and Freitas, Nando and Gordon, Neil},
booktitle = {Sequential Monte Carlo Methods in Practice},
doi = {10.1007/978-1-4757-3437-9_1},
pages = {3--14},
publisher = {Springer New York},
title = {{An Introduction to Sequential Monte Carlo Methods}},
url = {http://link.springer.com/10.1007/978-1-4757-3437-9{\_}1},
year = {2001}
}

@book{Beskos2014,
archivePrefix = {arXiv},
arxivId = {1103.3965},
author = {Beskos, Alexandros and Crisan, Dan and Jasra, Ajay},
booktitle = {Annals of Applied Probability},
doi = {10.1214/13-AAP951},
eprint = {1103.3965},
isbn = {1550001191},
issn = {10505164},
number = {4},
pages = {1396--1445},
title = {{On the stability of sequential Monte Carlo methods in high dimensions}},
volume = {24},
year = {2014}
}

@article{Green1995,
author = {Green, Peter J.},
doi = {10.1093/biomet/82.4.711},
isbn = {00063444},
issn = {00063444},
journal = {Biometrika},
number = {4},
pages = {711--732},
title = {{Reversible jump Markov chain monte carlo computation and Bayesian model determination}},
volume = {82},
year = {1995}
}

@article{Gordon1993a,
author = {Gordon, N. J. and Salmond, D. J. and Ewing, C. M.},
doi = {10.2514/3.21565},
issn = {07315090},
journal = {Guidance, Navigation and Control Conference, 1993},
number = {6},
pages = {11--21},
title = {{Bayesian state estimation for tracking and guidance using the bootstrap filter}},
volume = {18},
year = {1993}
}

@article{VanLeeuwen2009a,
author = {{Van Leeuwen}, Peter Jan},
doi = {10.1175/2009MWR2835.1},
issn = {00270644},
journal = {Monthly Weather Review},
number = {12},
pages = {4089--4114},
title = {{Particle filtering in geophysical systems}},
volume = {137},
year = {2009}
}

@book{Evensen2009,
author = {Evensen, Geir},
booktitle = {Vasa},
doi = {10.1007/978-3-642-03711-5},
isbn = {9783642037108},
issn = {14337851},
pages = {xxiii + 307},
title = {{Data Assimilation - The Ensemble Kalman Filter}},
url = {http://www.springer.com/book/978-3-642-03710-8{\%}5Cnhttp://medcontent.metapress.com/index/A65RM03P4874243N.pdf},
year = {2009}
}

@book{Doucet2001,
address = {New York, NY},
author = {Doucet, Arnaud and {De Freitas}, Nando and Gordon, Neil},
doi = {10.1007/978-1-4757-3437-9},
editor = {Doucet, Arnaud and Freitas, Nando and Gordon, Neil},
isbn = {978-1-4419-2887-0},
month = {apr},
publisher = {Springer New York},
title = {{Sequential Monte Carlo Methods in Practice}},
url = {http://link.springer.com/10.1007/978-1-4757-3437-9},
year = {2001}
}

@article{Gordon1993,
author = {Gordon, N. J. and Salmond, D. J. and Smith, A. F.M.},
doi = {10.1049/ip-f-2.1993.0015},
issn = {0956375X},
journal = {IEE Proceedings, Part F: Radar and Signal Processing},
number = {2},
pages = {107--113},
title = {{Novel approach to nonlinear/non-gaussian Bayesian state estimation}},
volume = {140},
year = {1993}
}

@article{Kantas2015,
archivePrefix = {arXiv},
arxivId = {1412.8695},
author = {Kantas, Nikolas and Doucet, Arnaud and Singh, Sumeetpal S. and Maciejowski, Jan and Chopin, Nicolas},
doi = {10.1214/14-STS511},
eprint = {1412.8695},
issn = {0883-4237},
journal = {Statistical Science},
number = {3},
pages = {328--351},
title = {{On Particle Methods for Parameter Estimation in State-Space Models}},
url = {http://projecteuclid.org/euclid.ss/1439220716},
volume = {30},
year = {2015}
}

@book{Brooks2011,
doi = {10.1201/b10905},
editor = {Brooks, Steve and Gelman, Andrew and Jones, Galin and Meng, Xiao-Li},
isbn = {978-1-4200-7941-8},
month = {may},
pages = {2011},
publisher = {Chapman and Hall/CRC},
series = {Chapman {\&} Hall/CRC Handbooks of Modern Statistical Methods},
title = {{Handbook of Markov Chain Monte Carlo}},
url = {http://www.crcnetbase.com/doi/abs/10.1201/b10905-6 http://www.crcnetbase.com/doi/book/10.1201/b10905 https://www.taylorfrancis.com/books/9781420079425},
volume = {20116022},
year = {2011}
}

@article{Vetra-Carvalho2018,
author = {Vetra-Carvalho, Sanita and van Leeuwen, Peter Jan and Nerger, Lars and Barth, Alexander and Altaf, M. Umer and Brasseur, Pierre and Kirchgessner, Paul and Beckers, Jean Marie},
doi = {10.1080/16000870.2018.1445364},
issn = {16000870},
journal = {Tellus, Series A: Dynamic Meteorology and Oceanography},
number = {1},
pages = {1--38},
publisher = {Taylor {\&} Francis},
title = {{State-of-the-art stochastic data assimilation methods for high-dimensional non-Gaussian problems}},
url = {https://doi.org/10.1080/16000870.2018.1445364},
volume = {70},
year = {2018}
}

@article{Surace2019,
archivePrefix = {arXiv},
arxivId = {1703.07879},
author = {Surace, Simone Carlo and Kutschireiter, Anna and Pfister, Jean Pascal},
doi = {10.1137/17M1125340},
eprint = {1703.07879},
issn = {00361445},
journal = {SIAM Review},
number = {1},
pages = {79--91},
title = {{How to avoid the curse of dimensionality: Scalability of particle filters with and without importance weights}},
volume = {61},
year = {2019}
}

@article{VanLeeuwen2019,
archivePrefix = {arXiv},
arxivId = {1807.10434},
author = {van Leeuwen, Peter Jan and K{\"{u}}nsch, Hans R. and Nerger, Lars and Potthast, Roland and Reich, Sebastian},
doi = {10.1002/qj.3551},
eprint = {1807.10434},
issn = {1477870X},
journal = {Quarterly Journal of the Royal Meteorological Society},
number = {723},
pages = {2335--2365},
title = {{Particle filters for high-dimensional geoscience applications: A review}},
volume = {145},
year = {2019}
}

@article{Li2014a,
author = {Li, Tiancheng and Sun, Shudong and Sattar, Tariq Pervez and Corchado, Juan Manuel},
doi = {10.1016/j.eswa.2013.12.031},
issn = {09574174},
journal = {Expert Systems with Applications},
number = {8},
pages = {3944--3954},
publisher = {Elsevier Ltd},
title = {{Fight sample degeneracy and impoverishment in particle filters: A review of intelligent approaches}},
url = {http://dx.doi.org/10.1016/j.eswa.2013.12.031},
volume = {41},
year = {2014}
}

@article{Somogyvari2019,
author = {Somogyv{\'{a}}ri, M{\'{a}}rk and Reich, Sebastian},
doi = {10.1007/s11004-019-09811-x},
issn = {1874-8961},
journal = {Mathematical Geosciences},
month = {may},
publisher = {Springer Berlin Heidelberg},
title = {{Convergence Tests for Transdimensional Markov Chains in Geoscience Imaging}},
url = {http://link.springer.com/10.1007/s11004-019-09811-x},
year = {2019}
}

@article{Snyder2008,
author = {Snyder, Chris and Bengtsson, Thomas and Bickel, Peter and Anderson, Jeff},
doi = {10.1175/2008MWR2529.1},
issn = {00270644},
journal = {Monthly Weather Review},
number = {12},
pages = {4629--4640},
title = {{Obstacles to high-dimensional particle filtering}},
volume = {136},
year = {2008}
}

@article{larocque2002particle,
  title={Particle filters for tracking an unknown number of sources},
  author={Larocque, J-R and Reilly, James P and Ng, William},
  journal={IEEE Transactions on Signal Processing},
  volume={50},
  number={12},
  pages={2926--2937},
  year={2002},
  publisher={IEEE}
}

\end{document}